\documentclass[structabstract]{aa}

\usepackage{times,amsmath,amssymb,txfonts,graphicx,aas_macros,natbib}
\usepackage{xspace,color,multirow,pbox}

\usepackage{relsize}

\newcommand\kms{\,\rm km\,s^{-1}}

\newcommand\muG{\,\mu\rm G}

\newcommand\kpc{\,\rm kpc}

\newcommand\perccm{\,\rm cm^{-3}}

\newcommand\rr{\hat{{\mathbf r}}}
\newcommand\zz{\hat{{\mathbf z}}}
\newcommand\vv{{\rm v}}
\newcommand\vvp{{\rm v'}}
\newcommand{\avg}[1]{\left<\right.\!#1\!\left.\right>}

\begin{document}

\title{Do magnetic fields influence gas rotation in galaxies?}

\author{D. Elstner \inst{1} , R. Beck \inst{2} \and O. Gressel\inst{3}}
\institute{Leibniz-Institut f\"ur Astrophysik Potsdam,
           An der Sternwarte 16, 14482 Potsdam, Germany;
           \email{delstner@aip.de}\\
           Max-Planck-Institut f\"ur Radioastronomie,
           Auf dem H\"ugel 69, 53121 Bonn, Germany;
           \email{rbeck@mpifr-bonn.mpg.de}\\
           Niels Bohr International Academy, Niels Bohr Institute,
           Blegdamsvej 17, 2100 Copenhagen \O, Denmark;
           \email{gressel@nbi.dk}
}

\date{Received ; accepted }

\abstract
    {}
    {We aim to estimate the contribution of the radial component of
      the Lorentz force to the gas rotation in several types of galaxies.}
    {Using typical parameters for the exponential scale of synchrotron
      emission and the scale length of HI gas, under the assumption
      of equipartition between the energies of cosmic rays and total magnetic fields,
      we derive the Lorentz force and compare it to the gravitational
      force in the radial component of the momentum equation.  We
      distinguish the different contributions between the large-scale
      and the small-scale turbulent fields by Reynolds averaging. We
      compare these findings with a dynamical dynamo model.}
    {We find a possible reduction of circular gas velocity in the 
      outermost parts and an increase inside a radius of four times the
      synchrotron scale length. Sufficiently localised radial
      reversals of the magnetic field may cause characteristic
      modulations in the gas rotation curve with typical amplitudes of
      $10-20\kms$.}
    {It is unlikely that the magnetic field contributes to the flat
      rotation in the outer parts of galaxies.  If anything, it will
      \emph{impede} the gravitationally supported rotation, demanding
      an even higher halo mass to explain the observed rotation
      profile. We speculate that this may have consequences for ram
      pressure stripping and the truncation of the stellar disc.}


    \titlerunning{Magnetic support of gas rotation}

    \maketitle


\section{Introduction}

The rotation curves observed in galaxies are generally explained
by a dominating contribution of dark matter \citep[see e.g. the review
  by][]{2001ARA&A..39..137S}, but the nature of dark matter is still
not known. 
Several ideas have been proposed in the literature that the
Lorentz force could support the rotation of the gas and hence reduce
the need for dark matter
\citep{2000FCPh...21....1B,2010ApJ...723L..44R,2012ApJ...755L..23R,%
  2012MNRAS.427.393J,2013MNRAS.433.2172S}. This however requires
Alfv\'en speeds of the order of $100\kms$, and moreover a large radial
extent of the magnetic field. Observed motions of stars or stellar clusters
also favour gravitation for the support of gas rotation \citep{1993MNRAS.261L..21P}.

The increased amount and quality of radio continuum data enables a
re-investigation of the role of magnetic fields for the rotation of
gas in galaxies.  From the radial scale lengths of synchrotron
emission and the gas density scale length, the Alfv\'en speed and its
radial profile can be computed. The contribution of magnetic forces to
the gas rotation can be estimated for various profiles of the radial
distribution of the magnetic field.

Magnetic fields in galaxies are mostly turbulent and are closely
related to star-forming activity \citep{2013A&A...552A..19T}. The
weaker regular fields show a coherent structure over kpc scales,
which may be due to a mean-field $\alpha\Omega$ dynamo
\citep{1996ARA&A..34..155B,2008AN....329..619G,2008A&A...486L..35G}.
There is observational evidence that regular fields have larger radial
scale lengths than turbulent fields because their amplification is not
restricted to energy input from star formation. The action of the
magneto-rotational instability (MRI) may be responsible for the field
amplification in the outer parts \citep{1999ApJ...511..660S}. The
field may appear more ordered because of a highly anti-parallel field
connected to the channel modes \citep{2009IAUS..259..467E}.  As a
consequence, the azimuthal and radial components of the regular field
may contribute to magnetic forces in the outer parts of a galaxy.

The general concept is presented in Sect.~2.
From radial profiles of the total and polarised synchrotron intensity
we derive the magnetic field strength of the turbulent and regular
magnetic components (Sect.~3). Together with gas density profiles we estimate
the influence of the Lorentz force on the gas rotation (Sect.~4). Moreover, we
use recent results from a \emph{hybrid} dynamo simulation for
a Milky Way-type galaxy \citep*{2013A&A...560A..93G} to test our
findings (Sect.~5). In this simulation, the classical $\alpha\Omega$ description
of the mean-field dynamo is complemented with a dynamically evolving
disc model. This treatment allows for the development of magnetic
instabilities, which occur on scales large enough not to be
affected immediately by turbulent diffusion. One outcome of the
simulation is a radial reversal in the magnetic field strength, which
we show to be associated with a local modulation of the galactic
rotation curve.


\section{The Lorentz force in galactic discs }

The radial component of the stationary momentum equation in
cylindrical geometry is
\begin{multline}
  \rho \left( \vv_r \partial_{r} \vv_r + \frac{\vv_\varphi}{r}
  \partial_\varphi \vv_r + \vv_z \partial_z \vv_r -
  \frac{\vv_\varphi^2}{r} \right) \\ = - \partial_r p + \rho g_r +
  j_\varphi B_z - j_z B_\varphi\,,
  \label{eq0}
\end{multline}
where the last two terms represent the Lorentz force, ${\,\mathbf
  j}\times{\mathbf B}$ and $\rho g_r$ the gravitational force.
Neglecting the radial velocity in the above
equation yields
\begin{equation}
  \vv_\varphi^2 = - r\,g_r + \frac{r}{\rho} \partial_r p -
  \frac{r}{\rho}\left(j_\varphi B_z - j_z B_\varphi\right)\,.
  \label{eq0a}
\end{equation}
In order to discuss the influence of the magnetic field on the
circular rotation we define the magnetic rotation velocity as
\begin{equation}
\vv_{\rm mag}^{ 2 }=\frac { 1 }{ \mu \rho } \left[ \left(
  \partial_{r} rB_{ \varphi } - \partial _{ \varphi }B_{ r } \right) B_{
    \varphi } - r \left( \partial _{ z }B_{ r }- \partial _{ r }B_{ z
  } \right) B_{ z } \right]\,,
\label{eq1}
\end{equation}
and the gravitational part, also including the small contribution of
gas pressure
\begin{equation}
  \vv_{\rm grav}^2= - r g_r + \frac{r}{\rho} \partial_r p\,.
  \label{eq1a}
\end{equation}
The gravitationally balanced rotation velocity is usually split into
contributions from stellar and gaseous disc, bulge and dark matter,
but we will not discuss the details of the individual gravitational
effects here.  The total circular velocity of the gas is given by
\begin{equation}
  \vv_{ \varphi  }=\sqrt{\vv_{\rm grav}^2+\vv_{\rm mag}^2}.
  \label{eq3}
\end{equation}

We first consider a large-scale axisymmetric field with a
negligible vertical field component, as typical for the
regular magnetic field in the midplane of nearby galaxies
\citep{2013pss5.book..641B}.  Then only the first term of the right-hand side of
Eq.~(\ref{eq1}) has to be considered.  For a power-law distribution of
the magnetic field strength $ B_{ \varphi }=B_{0}r^{ n }$, we get
\begin{equation}
  \vv_{\rm mag}^{ 2 }(r) = \frac { B_{ 0 }^{ 2 } }{ \mu \rho }
  \left( \left( n+1 \right) r^{ 2n } \right) = (n+1)\,\vv_{\rm A}^2(r)
  \label{eq2}
\end{equation}
with the Alfv\'en speed $\vv_{\rm A} \equiv B_{\varphi}\,(\mu
\rho)^{-1/2}$.

For a radial profile of the azimuthal  field decaying slower than
$r^{-1}$, the Lorentz force acts inwards and needs to be balanced
by the centrifugal force of the gas rotation, hence reducing the need
for gravitational pull from dark matter.  For a decay profile of the
magnetic field steeper than $r^{-1}$ the right hand side of
Eq.~(\ref{eq2}) becomes negative, leading to a slower rotation than
predicted by the balance of the centrifugal force with the
gravitational force only, with the consequence that even more dark
matter is needed than in the non-magnetic case.

The singularity of the $r^{-1}$ profile at the origin can be avoided
by an ansatz
\begin{equation}
  B_{ \varphi  }=B_{0} \left( 1+\frac{r}{r_{\rm m}} \right)^{-1}
  \label{eq3aa}
\end{equation}
as used by \citet{2012ApJ...755L..23R}, leading to a magnetic velocity
\begin{equation}
  \vv_{\rm mag}^{ 2 }(r) = \left(1- \frac{r}{r_{\rm m}+r} \right) {\rm
    v}_{\rm A}^2(r)\,.
  \label{eq3ab}
\end{equation}
This more or less artificial approach always yields an inward directed
force, because it is always flatter than a $1/r$ profile, which is
unlikely because it corresponds to a vertical current with a single
sign through the whole galactic disc.  Dynamo-excited magnetic fields
never show such configurations \citep{2013A&A...560A..93G}.

The  Alfv\'en velocity is an upper limit for the change of the rotation by the magnetic force
for the above defined profiles.
Typical values of 30$\kms$ in the thin disc may affect the circular
velocity of 200$\kms$ only by 1\%.  The situation may be different in
the very outer disc or in the halo where the Alfv\'en speed may be
150$\kms$ or higher because of the density decrease in gas density.
However, if the generally observed relation $B \propto \rho^{0.5}$
\citep{2010ApJ...725..466C,1997A&A...320...54N}
also holds for the
large-scale field in the outer disc, a constant Alfv\'en speed follows
(see Sect.~\ref{obs}).

In accordance with exponential profiles of synchrotron emission, we assume
an exponential profile for the regular magnetic field with scale
length $r_{\rm m}$ (see Sect.~\ref{obs}).  Here we find
\begin{equation}
  \vv_{\rm mag}^{ 2 }(r) = \frac { B_{ 0 }^{ 2 } }{ \mu \rho }
  \left(1-r/r_{\rm m} \right) \exp(-2r/r_{\rm m})
  = \left(1-r/r_{\rm m} \right){\rm v}_{\rm A}^2(r)\,.
  \label{eq3a}
\end{equation}
Now the magnetic force is directed inwards (positive $\vv_{mag}^2$)
only in the inner part $r < r_{\rm m}$.  In the outer parts it acts
outwards against the gravitational force. We note that the absolute
value of the magnetic force increases already for a constant Alfv\'en speed.  In this
case the vertical current associated with the azimuthal magnetic field
changes its sign at the scale length.  The situation is quite similar
for a Gaussian profile,
\begin{eqnarray}
  \vv_{\rm mag}^{ 2 }(r) &=& \frac { B_{ 0 }^{ 2 } }{ \mu \rho }
  \left(1-2 (r/r_{\rm m})^2 \right) \exp(-2(r/r_{\rm m})^2) \nonumber \\ &=&
  \left(1-2 (r/r_{\rm m})^2 \right)\vv_{\rm A}^2(r) \, ,
  \label{eq3b}
\end{eqnarray}
where the sign of the magnetic force changes at $r=r_{\rm m}/\sqrt{2}$.

We now consider magnetic fields also including  radial and vertical
fields. Rewriting Eq.~(\ref{eq1}) using ${\rm div}(B)=0$, we obtain
\begin{eqnarray}
  \vv_{\rm mag}^{ 2 }&=&\frac { 1 }{ \mu \rho }
  \left( B_{\varphi}\partial _{r} rB_{\varphi}
       - B_{ r }\partial _{r}rB_{r} + rB_{z}\partial_{r}B_{z}
  \right.\nonumber \\
  & & \left. - \partial_{\varphi} (B_{r}B_{\varphi}) -
  r\partial_{z}(B_{r}B_{z})\right)\,.
  \label{eq40}
\end{eqnarray}
Because we are interested in a mean radial dependence,
we apply the
average $\avg{\,} \equiv \frac{1}{4\pi h} \int_{0}^{2\pi}{
  \int_{-h}^{h}{ dz\, d\varphi } }$ to Eq.~(\ref{eq40}) to yield
\begin{eqnarray}
  \avg{\rho \vv_{\rm mag}^{ 2 }} &=&\frac { 1 }{ \mu }
      \left[\,
        \left( \avg{B_{\varphi}^2}
        + \frac{r}{2}\partial _{ r } \avg{B_{\varphi}^2} \right)
        - \left( \avg{B_{r}^2}
             + \frac{r}{2}\partial_{r} \avg{B_{r}^2} \right) \right.
        \nonumber \\ &
    &- \frac{r}{4\pi h} \int_{0}^{2\pi}{ \left(\, B_r(h) B_z(h) - B_r(-h)
      B_z(-h)\,\right) d\varphi} \nonumber \\ & &
        \left. + \frac{r}{2}\partial_{r}\avg{B_{z}^2}
        \,\right]\,.
  \label{eq4a}
\end{eqnarray}
The term in the second line vanishes for periodic boundary
conditions, which may be true for the turbulent field.  For a
symmetric or an antisymmetric field with respect to the galaxy's plane,
we have $ - 2 r
\int_{0}^{2\pi}{B_r(h) B_z(h)d\varphi}$ which always contributes with
a negative sign for an X-shaped structure as is seen in many
edge-on galaxies \citep{2014arXiv1401.1317K}.

We note that $B_z=0$ implies $B_r \propto r^{-1}$ in the axisymmetric case and
Eq.~(\ref{eq4a}) agrees with Eq.~(\ref{eq1}).  For a given radial
distribution of an axisymmetric $B_r$, the radial distribution of the
vertical flux from the disc is fixed by ${\rm div} B= 0$.
Consequently, we can set the vertical field at the disc surface
\begin{equation}
  B_z(h)=-B_z(-h)=-\frac{h}{r}
\partial_r \left(\,r \avg{B_r}\,\right)\,.
\end{equation}
Here we consider only a symmetric field with respect to the plane. For an antisymmetric field $B_z$ is an independent
variable.

The left-hand side of Equation~(\ref{eq4a})
refers to the total magnetic energy of the three
field components, which can be estimated from the synchrotron emission
under the assumption of equipartition with cosmic rays.  For the cross
correlation term at least the sign is known.  A flat radial profile of
the azimuthal field contributes to an inward-directed Lorentz
force, but radial magnetic fields are not negligible in galaxies.
For a constant pitch angle, the magnetic force from the  azimuthal field
will be reduced by the contribution of the radial component.  For
pitch angles at about 20$^\circ$, the force will be reduced by 14\% ;
for an angle of 35$^\circ$, the effect will amount to about 50\%.
This  effect only occurs for anti-parallel fields, in contrast to a
uniform regular field, where the radial component has no influence.
The turbulent magnetic field also contributes to the Lorentz force as
a pressure term to the momentum equation for isotropic fields
(cf. Eq.~\ref{eq5}).  For anisotropic turbulence (i.e. 
$\avg{B_{\varphi}^2} \not= \avg{B_{r}^2}$), we may also get tension
effects similar to the regular fields.

\noindent Defining
\begin{eqnarray}
  \vv_{M}^{ 2 } & = &\frac { 1 }{ \mu \avg{\rho}}  \left[\;
    \left( \avg{B_{\varphi}^2}
    + \frac{r}{2}\partial _{ r }\avg{B_{\varphi}^2} \right) \right.
    - \left( \avg{B_{r}^2} + \frac{r}{2}\partial _{ r }\avg{B_{r}^2} \right)
    \nonumber \\
    & &-  \frac{r}{4\pi h} \int_{0}^{2\pi}{
      \left(\, B_r(h) B_z(h) - B_r(-h) B_z(-h)
      \,\right)\,d\varphi} \nonumber \\
    & & \left. + \frac{r}{2}\partial _{r}\avg{B_{z}^2} \;\right]
  \label{eq5}
\end{eqnarray}
and the rotation velocity $\vv_G$ balancing the gravitational force we
get the circular velocity
\begin{equation}
  \!\avg{\vv} = \sqrt{\frac{\avg{\rho'\vvp}^2}{\avg{\rho}^2}
              + \vv_G^2 + \vv_M^2 - \avg{\vvp^2}
              - \frac{\avg{\rho'\vvp^2}}{\avg{\rho}}}
              - \frac{\avg{\rho'\vvp}}{\avg{\rho}}
  \label{eq6}
\end{equation}
by applying Reynolds rules to the term $\avg{\rho \vv^2}$. 
Here we neglect all effects of velocity and density fluctuations
including the asymmetric drift term $\avg{\vvp^2}$. We finally obtain
the circular velocity supported by gravitation and magnetic fields,
\begin{equation}
  \avg{\vv} = \sqrt{\vv_G^2 + \vv_{M}^2}\,.
  \label{eq7}
\end{equation}
It is easy to see that $\vv_M$ coincides with the magnetic velocity
$\vv_{\rm mag}$ defined in Eq.~(\ref{eq1}). In contrast,
Equation~(\ref{eq5}) serves as an alternative formulation for the
magnetic velocity with energy densities of the total magnetic field
components, which allows a better interpretation of the observed data.


\section{Observed radial profiles of magnetic fields and gas in galaxies}
\label{obs}

The radial profiles of the total and linearly polarised radio synchrotron
intensities in nearby galaxies can be well fitted by exponential distributions.
The exponential scale lengths for grand-design spirals are
typically $5\kpc$ for the polarised and about $4\kpc$ for the total
synchrotron emission, for instance in NGC~6946 \citep[see][]{2002A&A...388....7W}.

The scale length of the total magnetic field $B_{\rm tot}$ follows from that of
the total synchrotron emission, assuming equipartition between the total magnetic
and cosmic ray energy densities (see below).
The scale length of the regular magnetic field $B_{\rm reg}$ follows from that of
the total synchrotron emission and the degree of linear polarisation. As the degree of
polarisation generally increases with increasing radius, the scale length of $B_{\rm reg}$
is larger than that of $B_{\rm tot}$, but has a larger uncertainty.
We note that the \emph{anisotropic} field component also
contributes to the polarised intensity, but it would affect the
magnetic velocity in a similar way to the turbulent field, where the
effect of the azimuthal field is reduced for non-zero pitch angles 
(cf. Eq.~(\ref{eq5})).

Under the equipartition assumption and assuming a radio synchrotron spectral index of about $-1$,
\begin{equation}
  {\rm PI} \;\propto\; N_{\rm CR} B_{\rm reg}^2
     \;\propto\; B_{\rm tot}^2 B_{\rm reg}^2
     \;{\propto}\; B_{\rm reg}^4\,,
\end{equation}
and hence $I \propto N_{\rm CR} B_{\rm tot}^2 \propto B_{\rm tot}^4$
\citep{2007A&A...470..539B}, the scale length of the {\bf total} magnetic field is
about four times larger than that of the {\bf total} synchrotron emission.
The profile of the regular field is described as $B_{\rm reg} = B_{\rm
  reg,0} \exp(-r/R_{B\,{\rm reg}})$ and that of the turbulent field
$B_{\rm tur} = B_{\rm tur,0} \exp(-r/R_{B\,{\rm tur}})$.  The observed
synchrotron scale lengths correspond to a typical scale length of the
regular field of $R_{B{\rm reg}}\simeq 20\kpc$ and a scale length of
the total field of $R_{B{\rm tot}}\simeq 16\kpc$.  These values vary
with the size of the galaxies and do strongly scatter. Synchrotron
scale lengths for M83 \citep{1987MNRAS.227..887H,1993A&A...274..687N}
and the Milky Way are about $2.5\kpc$ and may reach $7\kpc$ for
M101 \citep{1990ASSL..160..141H}
and NGC~6753 \citep{1987MNRAS.227..887H}.

Mean-field $\alpha\Omega$ dynamo models also show an exponential radial profile of
the magnetic field, but mostly with shorter scale lengths of the order
of $5\kpc$ for grand design spirals \citep{2013A&A...560A..93G}.  In
the cosmological simulations of \cite{2012MNRAS.422.2152B}, the field
strength in  protogalactic halos approximately follows a $1/r$ profile.

The total field has two components: regular and turbulent. The
turbulent field may be anisotropic, where the extreme case of a
vanishing radial and vertical component would still contribute to
polarised emission.  Hence, the exponential scale length of the total
field is a weighted mean of the individual field components.  As the
turbulent field dominates
\citep{2007A&A...470..539B,2013A&A...552A..19T}, we assume a typical
scale length of the turbulent field of $R_{B{\rm tur}} \simeq 15\kpc$
for spiral galaxies.

The synchrotron scale lengths of dwarf galaxies are smaller than in
spirals, for instance $0.6\kpc$ in NGC~4214
\citep{2011ApJ...736..139K}.  The corresponding scale length of the
total and turbulent fields is about $3\kpc$. Polarised emission has
been detected only for dwarf galaxies with a high star formation rate,
like NGC~4449 \citep{2000A&A...355..128C}, for which we assume a typical scale
length of the regular field of $R_{B{\rm reg}}\simeq 4\kpc$.

The strength of the turbulent field $B_{\rm tur}$ depends on the
star formation rate. At $r=0$, $B_{\rm tur,0}\simeq 20\muG$ is typical
for a grand-design spiral like NGC~6946 \citep{2007A&A...470..539B},
while for galaxies with lower star formation rates we assume
$B_{\rm tur,0}\simeq 10\muG$. The strongly star-forming dwarf galaxy
NGC~4449 has $B_{\rm tur,0}\simeq 15\muG$ \citep{2000A&A...355..128C},
while weakly star-forming dwarfs reveal $B_{\rm tur,0}\simeq 3\muG$
\citep{2011A&A...529A..94C}.

The strength of the regular field $B_{\rm reg}$ does not show a clear
dependence on a single parameter of the host galaxy.  We assume a
ratio of $B_{\rm reg,0} / B_{\rm tur,0} = 0.4$ for spiral galaxies,
corresponding to a mean degree of polarisation of $p = p_i \, B_{\rm
  reg,0}^2 / (B_{\rm reg,0}^2 + B_{\rm tur,0}^2) \simeq 10\%$. This
value increases with radius $R$ because of the larger scale length of
$B_{\rm reg}$. Dwarf galaxies have weaker regular fields even if the
star formation rate is high ($B_{\rm reg,0}\simeq 3\muG$). No regular
fields have been detected so far in dwarf galaxies with a low
star formation rate.

Typical parameters of the magnetic field strengths near the centre
($r=0$) and the magnetic scale lengths are given in
Table~\ref{tab1} for different galaxy types. We also include values
for the Milky Way \citep{2008A&A...487..951K,1998ApJ...497..759F,%
  1985ApJ...295..422C,1978A&A....63....7B}, NGC~6946
\citep{2002A&A...388....7W,2008A&A...490..555B} and M31 \citep{1982A&A...105..192B}.
 In the last column we give estimates for the Alfv\'en velocity of the total field near the centre.


\begin{table*}[tdp]
\caption{Observational parameters for typical spiral and dwarf
  galaxies (see text for details). The Milky Way, NGC~6946 and M31 are included
  because these are the cases with the most complete observations.}
\begin{tabular}{lcccccccc}
\hline
Galaxy              & $\!\!B_{\rm reg,0\!}$ & $\!B_{\rm tur,0\!}$
                    & $\!\!R_{B\,{\rm reg}}\!$ & $\!R_{B\,{\rm tur}}\!$
                    & $R_{\rm gas}$ & $\!R_\Omega\!$ & $\Omega_0$ & $V_a$\\[1pt]
                    & $\![\mu{\rm G}]\!$ & $\![\mu{\rm G}]\!$
                    &$\![\rm kpc]\!$ & $\![\rm kpc]\!$ & $\![\rm kpc]\!$
                    & $\!\![\rm kpc]\!\!$ & $\![{\rm Gyr}^{-1}]\!$  & km/s \\[3pt]
\hline
\\[-6pt]
Spiral, high~SF     &  8 &  20 &  20 &  15 &  5-10   &  1 & $\!\!\!\!$200-250$\!$ & 12 \\
Spiral, low~SF      &  4 &  10 &  20 &  15 &  5-10   &  1  & $\!\!\!\!$200-250$\!$ & 6 \\
Flocculent          &  4 &  10 &  20 &  15 &  5-10   &  3  &  50-70  & 6 \\
Dwarf,  high~SF     &  3 &  15 &   4 &   3 &   5   &  1  &  40-70  & 7 \\
Dwarf,  low~SF      &  0 &   5 &   0 &   3 &   5   &  1  &  10-30 & 2 \\
Large, stripped$^a$ &  8 &  20 &  20 &  15 &   2   &  1  &   200  & 45  \\
Milky Way           &  5 &  11 &  10 &  10 &   3.1 &  .5 &   400 & 6.6  \\
NGC~6946    &  8 &  20 &  20 &  15 &   8.7 &  1      &   180 & 12   \\
M31, $R\!>\!10\kpc$ &  5 &   6 &  20 &  15  &  10   &  1  &   200 & 5  \\[3pt]
\hline
\end{tabular}\\[3pt]
$^a$ central gas density, $\rho_{\rm gas}=1\perccm$, otherwise
$\rho_{\rm gas}=15\perccm$.\vspace{-1.5ex}
\label{tab1}
\end{table*}%


All non-magnetic effects like gravitational forces (stellar disc, gas
disc, dark matter, etc.) or thermal and turbulent gas pressure are
represented by the circular velocity which we prescribe by a Brandt
law
\begin{equation}
  \vv_{\rm grav}(r) = \Omega_0 r/ (1+(r/R_\Omega)^2)^{0.5}
\end{equation}
with the inner circular frequency $\Omega_0$ and the turnover radius
$R_\Omega$ where the rotation becomes flat. This simple approximation 
does not really reflect the universal rotation law of galaxies \citep{2007MNRAS.378...41S}, 
but will be sufficient for a comparison with the magnetic influence on the rotation.  

The ionisation fraction in the neutral gas component is still high
enough to treat the total gas as a single fluid.
\cite{2012ApJ...756..183B} found a universal exponential
radial scaling of the total gas surface density for nearby,
non-interacting disc galaxies, with a scale length of about $R_{\rm
  gas} \simeq 0.6 \, r_{25} \simeq 10\kpc$.  Furthermore we add a constant
value of $10^{-29}\,{\rm g\,cm}^{-3}$ for the intergalactic medium.
Typical parameters of the scale length $R_{\rm gas}$ are given in
Table~\ref{tab1} for different galaxy types. With the exception of
gas-stripped galaxies with a much reduced central density of
$\rho_{\rm gas}=1\perccm$, the density is $\rho_{\rm gas}=15\perccm$
at {\bf $r=0$}. We include the case of a spiral galaxy for which the gas has
been stripped by tidal interaction, but the magnetic field is still
unchanged.  Such a scenario may occur in the early phase of
interaction \citep{2012A&A...537A.143V}.

The scale length of the total gas density is in many cases similar to
half of the scale length for the regular field, in accordance with a
constant Alfv\'en velocity, but again with a  large scatter. The
most extreme deviation could be M101 with $28\kpc$ scale length for the
total field and $7\kpc$ for the gas scale length
\citep{1990ASSL..160..141H,2010AJ....140.1194B}.
Here the Alfv\'en velocity is exponentially but slowly increasing with a huge scale length
of $28\kpc$. So far there is no evidence that the magnetic field strength follows such a law
even at distances of more than $28\kpc$.
 For M83 we have
$R_{\rm gas}=7\,{\rm kpc}$ and $R_{B\,{\rm total}}=10\kpc$
\citep{2010AJ....140.1194B,1993A&A...274..687N}.

There is not always a single scale length, in particular for interacting
galaxies, such as M51, with about $R_{\rm gas}=5.5\kpc$
and $R_{B\,{\rm total}}=18\kpc$ (derived from radio data at 151\,MHz, assuming equipartition)
between $2-10\kpc$ radius; these values change in the outer disc to become about $R_{\rm gas}=2\kpc$
and $R_{B\,{\rm total}} = 8\kpc$ in the range $10-18\kpc$
\citep{2010AJ....140.1194B,2014arXiv1407.1312M}. 
Here we have the case for a slowly growing  Alfv\'en velocity reaching about $100\kms$ at a distance of $40\kpc$,
but the small exponential scale in the outer part allows only an outward directed Lorentz force.   
\citet{2013A&A...552A..19T} found $B_{tot}\propto\Sigma_{gas}^{0.23\pm0.01}$
in NGC~6946 (where $\Sigma_{gas}$ is the gas surface density), based on
many small regions within the bright star-forming regions of the galaxy.
The magnetic field distribution was derived via the equipartition assumption from that
of the synchrotron intensity, which is known to be smoother than that of the gas,
due to diffusion of cosmic-ray electrons \citep{2013MNRAS.435.1598B}.
Hence, the exponent of the above relation should be regarded as a lower limit and does not
question our assumption of a constant Alfv\'en velocity on large scales.

A scaling of $B_{reg}\propto\rho^{0.5}$ is also
in agreement with an equipartition field strength for the more
or less constant velocity dispersion.  For the typical field strength
of several $\mu$G and a density of $1\perccm$, the Alfv\'en velocity
is of the order of $10-20\kms$, which is too small to cause a significant
change of a $200\kms$ circular velocity.

The estimates for the
density are derived from observations of surface densities.  For a
constant vertical scale height of the gas disc, the scale length is
the same for the midplane densities.  For a flaring disc such as in
the Milky Way, the scale length for the midplane density will be
smaller.  In the case that the magnetic field extends with the same scale
length outside the optical radius (which is still not observed) the
Lorentz force would not be negligible anymore.

Grand-design spiral galaxies usually have a regular field strengths between
$4\muG$ and $8\muG$ for low and high star formation rates,
respectively \citep{2010ASPC..438..197F}.
The exponential scale length for the regular magnetic
field turns out to be twice the scale length of the HI density, which
means a constant Alfv\'en speed of around $10\kms$ up to the outer
parts of the gaseous disc.  This value can be lower in the central
region because of high molecular gas density.  Therefore the magnetic
energy is roughly in equipartition with the turbulent energy of the
gas in the midplane.  With such low values of the Alfv\'en velocity,
we exclude any influence of the magnetic field on the gas rotation
independent of the radial magnetic field distribution.  A similar
situation is found for flocculent galaxies.  For extensively
star-forming dwarf galaxies, the magnetic scale is only slightly
larger than the density scale length, which leads to a radially
decreasing Alfv\'en velocity of at most 10\% of the rotational
velocity.


\section{Effect on rotation for idealised radial profiles}

\begin{figure}
  \centering
  \includegraphics[width=\columnwidth]{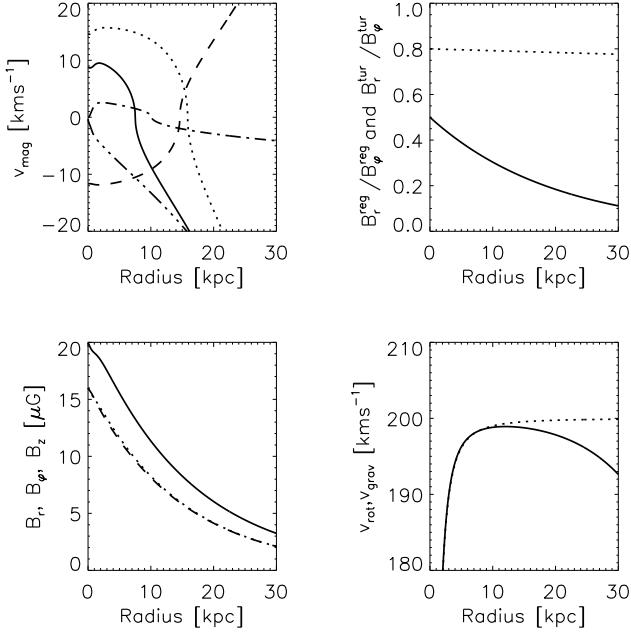}
  \caption{Typical spiral including a turbulent field.  In the upper-left panel 
  the contributions to the magnetic velocity (solid) are
    from the total field of $B_\varphi$ (dotted), $B_r$ (dashed),
    $B_z$ (dash dot dot), and the $B_r B_z$ term (dash dot) of the
    regular field only.  Upper-right panel shows the radial dependence
    of the pitch angle for the regular field (solid) and the
    anisotropy $q_{\rm iso}$ of the turbulent field (dotted).  
    Lower-left panel: radial profiles of total magnetic field components
    $B_\varphi$ (solid), $B_r$ (dotted), and $B_z$ (dashed)
     (the last two components are almost identical). 
     Lower-right panel: rotation modified by the Lorentz force (solid) and
    from the gravitation only (dotted).  }
  \label{fig1}
\end{figure}

We model the total magnetic field as the sum of the regular and
turbulent field. Because of the unknown contribution of the
anisotropic turbulent field to the polarised emission, we model that
part of the field as a regular field only.  We assume instead some
anisotropy for the turbulent field.  Prescribing the radial profile of
the regular $B_\varphi$ and the radial profile of the pitch angle
fixes the components $B_r$ and $B_z$ for a symmetric regular field.  For the
turbulent field we prescribe the radial profile of $B_\varphi^{\rm
  tur}$ and the degree of anisotropy ( $B_r^{\rm tur}=B_z^{\rm tur}=
q_{\rm iso} B_\varphi^{\rm tur}$). The third term of the right-hand side in
Eq.~(\ref{eq5}) vanishes for the turbulent field. Under the
assumptions for the radial profiles of density, regular, and turbulent
magnetic field for a typical spiral (cf. Tab.~\ref{tab1}) we plot the
circular velocity modified by the magnetic field in
Fig.~\ref{fig1}. The reduction of the magnetic velocity outside
$20\kpc$  is mainly caused by the pressure of the vertical component of
the turbulent field. The anisotropy of the turbulent field appears to
be negligible. For an antisymmetric regular field the result is
similar for a negligible magnetic flux through the disc.

\subsection{Large spirals}

Some large galaxies (like M51) have a truncated gas disc, where a much
smaller scale of the density is observed in the outer disc. The corresponding
truncation in the radio continuum disc may be ascribed to a lack of star formation
and hence of cosmic-ray electrons in the outer disc,
while the magnetic field may extend beyond the truncation radius with a constant
scale length. In such a case,
the Lorentz force may be influential in the outer parts.
Neglecting the turbulent field leads to an increase of the rotation
velocity by about $20\kms$ at $10\kpc$, where the Alfv\'en speed is
$130\kms$.  The situation changes if we also consider the turbulent
magnetic field acting as a pressure term only. Now the outward
pressure force would be in equilibrium with the gravitation at a
radius of $10\kpc$, but our assumption of equipartition between magnetic
and cosmic-ray energies gives an additional pressure
truncating the gaseous disc at a smaller radius.
Assuming a polytropic index of
$\gamma=4/3$ for the cosmic-ray gas, the cosmic-ray pressure is
$P_{\rm CR}=(1/3) P_{\rm mag}$, which reduces the truncation to 9kpc.

\begin{figure}
  \centering
  \includegraphics[width=\columnwidth]{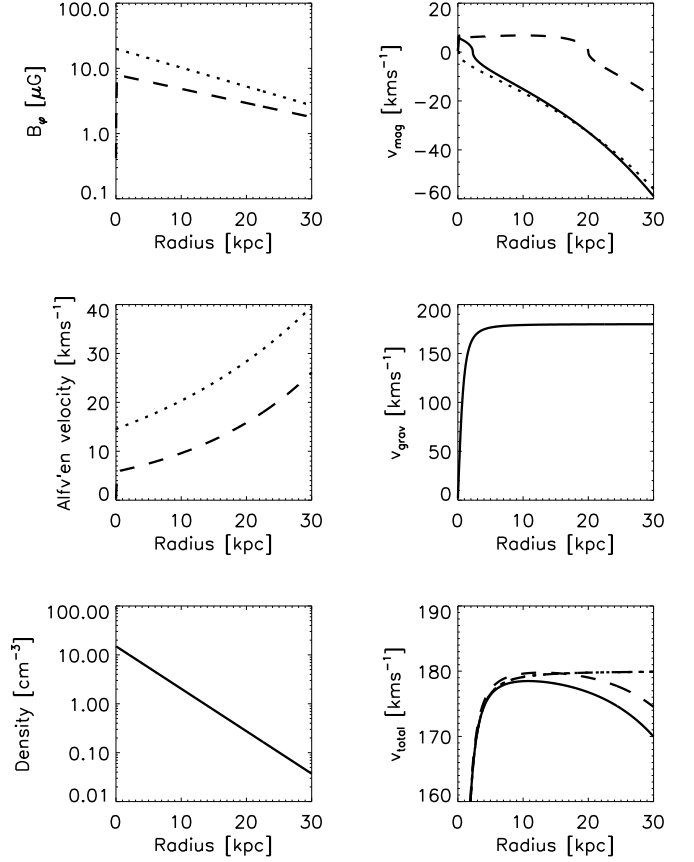}
  \caption{NGC~6946: Dotted lines in the upper panels denote the
    turbulent magnetic field and dashed lines the regular field.  The
    magnetic velocity $\vv_{\rm mag}$ for the total field (solid line) in
    the upper right panel.  In the lower right panel the resulting
    total circular velocity is plotted (solid). The dashed line  indicates
    the rotation velocity neglecting the turbulent field.}
  \label{fig2}
\end{figure}

We now consider two more or less constrained cases for typical spirals.
For NGC~6946 all parameters can be taken from 
\cite{2002A&A...388....7W,2007A&A...470..539B}; and \cite{2008A&A...490..555B} , 
which will be an example of a large spiral with moderate star formation.
Despite the exponential growth of the Alfv\'en speed in
Figure~\ref{fig2} we see only a weak reduction of the rotation velocity as a result
of the Lorentz force. This effect stems from the turbulent magnetic
field, which is included here only as a pressure term.
This galaxy is  one of the extreme cases were the magnetic energy is larger than the turbulent kinetic energy. 

\subsection{The Milky Way}

\begin{figure}
  \centering
  \includegraphics[width=\columnwidth]{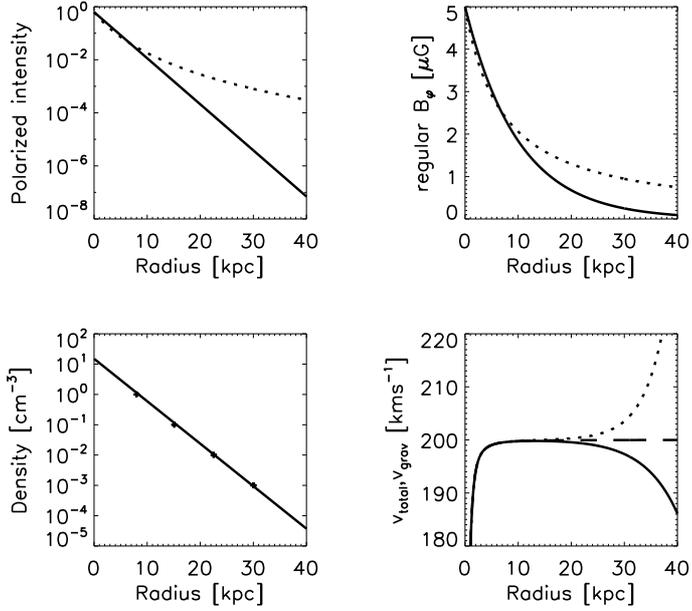}
  \caption{Milky Way: Circular velocity for a distribution of the
    regular magnetic field with an exponential profile (solid),  compared to the profile given by
    Eq.~(\ref{eq3aa}). The exponential scale length is $10\kpc$, and
    $r_{\rm m}=7\kpc$. One cannot distinguish the two
    profile shapes from a measurement of the (normalised) polarised
    intensity in the inner $10\kpc$. The density distribution of
    \cite{2008A&A...487..951K} is marked with crosses.}
  \label{mw1}
\end{figure}

Observations of the radial profile of the polarised intensity for the
Milky Way are less constrained than observations for external
galaxies. Estimates for the radial magnetic profile often rely on
models for the magnetic field
\citep{2008A&A...477..573S,2000ApJ...537..763S}. The scale length
of the magnetic field is nearly three times as big as the scale
length of the gas density, which gives an increase of the Alfv\'en velocity
(cf. Table~\ref{tab1}).  Here we compare the two basic assumptions for
the radial profile of the magnetic field.  The first model relies on
the assumption of equipartition between magnetic field and cosmic ray
energies. The density profile is taken from
\cite{2008A&A...487..951K}. For the second model, we used a magnetic
field profile according to Eq.~(\ref{eq3aa}), approximating a $1/r$
profile at infinity.  Assuming equipartition, this would lead to clear
contradiction with an exponential law for the polarised emission. On the other hand,
for an exponential cosmic ray distribution $\rho_{cr}=\exp(-r/r_{cr})$
with a scale length of $r_{cr}=6\kpc$, a nearly exponential
distribution of the polarised intensity is obtained. Figure \ref{mw1} shows the magnetic disturbance 
of the rotation for both assumptions. This leads to an increase in gas rotation
for the model of \cite{2012ApJ...755L..23R}, but a decrease in the
exponential magnetic field distribution. However, the effect is still
below $10\kms$ at $20\kpc$ radius in agreement with
\cite{2013MNRAS.433.2172S}, who used an exponential law.


\subsection{Field reversals}

\begin{figure}
  \centering
  \includegraphics[width=\columnwidth]{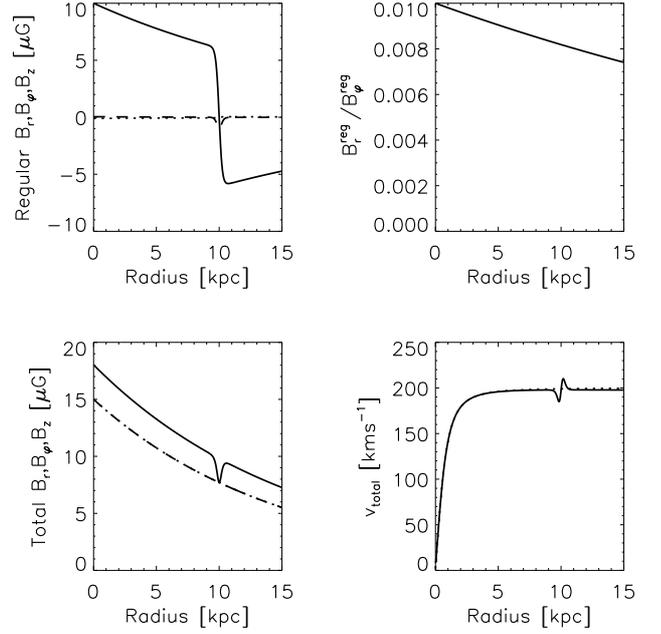}
  \caption{Milky Way: Reversal within 1\,kpc and a small pitch angle.
    Upper-left: regular $B\varphi$ (solid), $B_r$ (dotted), $B_z$
    (dashed); lower-left: total $B\varphi$ (solid), $B_r$ (dotted),
    $B_z$ (dashed); upper-right: regular $B_r/B_\varphi$ (solid);
    lower-right:  circular velocity with magnetic force (solid),
    gravitation only (dotted).}
  \label{rev1}
\end{figure}

\begin{figure}
  \centering
  \includegraphics[width=\columnwidth]{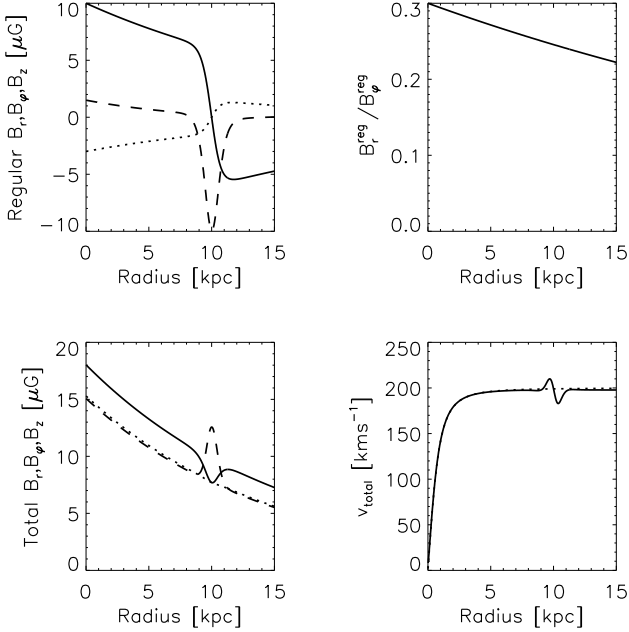}
  \caption{Milky Way: Reversal within 3\,kpc and a large pitch angle.
    Upper left: regular $B\varphi$ (solid) $B_r$ (dotted) $B_z$
    (dashed), lower left: total $B\varphi$ (solid) $B_r$ (dotted)
    $B_z$ (dashed), upper right: regular $B_r/B_\varphi$ (solid),
    lower right:  circular velocity with magnetic force (solid),
    gravitation only (dotted).}
  \label{rev2}
\end{figure}

For the Milky Way, a reversal of the regular magnetic field is a
possible interpretation of the data \citep[see][for example]{2011ApJ...728...97V}.
For the parameters of a typical
spiral (cf. Table~\ref{tab1}) with a low star formation rate, we
modelled a reversal of the regular field.  A sharp field reversal on a
length scale below $1\kpc$ can lead to strong modulations in the rotation
curve. For a dominant {\bf azimuthal} field (i.e., a small pitch angle), the
effect leads first to a reduction and then to an increase of the
circular velocity (cf. Fig.~\ref{rev1}) when moving radially
outwards. For larger pitch angles, the reversal of an axisymmetric
field also implies a strong vertical field, which results in a local
increase of the magnetic pressure. As a consequence, the deviation
from the unperturbed rotation curve starts with super-rotation and
ends with a weaker rotation (cf. Fig.~\ref{rev2}) when moving from
smaller to larger radii.  For a broader transition zone, the effect
becomes negligible.

\section{Dynamo model}

In this section, we aim to confront the theoretical expectations
derived above with a full-blown global simulation of a Milky Way-type
galaxy \citep*{2013A&A...560A..93G}, where we modelled a mean-field
dynamo in a dynamical fashion by additionally solving the momentum
equation in a prescribed static gravitational potential. In this
model, we found the action of the supernova-driven
$\alpha\Omega$~dynamo to be limited to the inner $10-15\kpc$. In the
outer part, the magnetic field was maintained by the magnetorotational
instability. The final magnetic field had a radial exponential scale
length of about $10\kpc$ and a modestly varying Alfv\'en speed of
$10-30\kms$.  Consistent with our preceding estimates, we found only a
very weak influence of the Lorentz force on the rotational velocity.

\begin{figure}
  \includegraphics[width=\columnwidth]{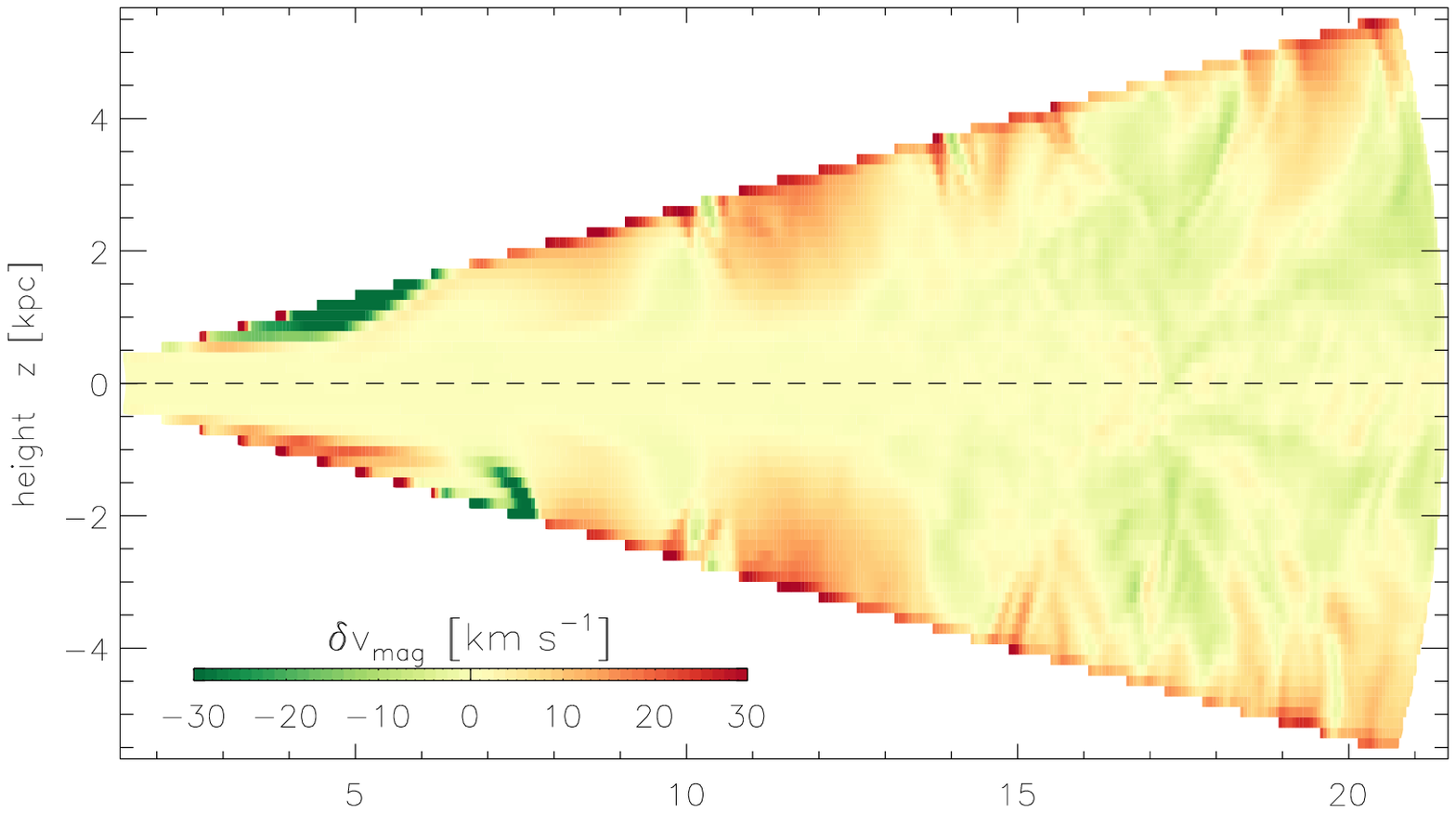}\\[6pt]
  \includegraphics[width=\columnwidth]{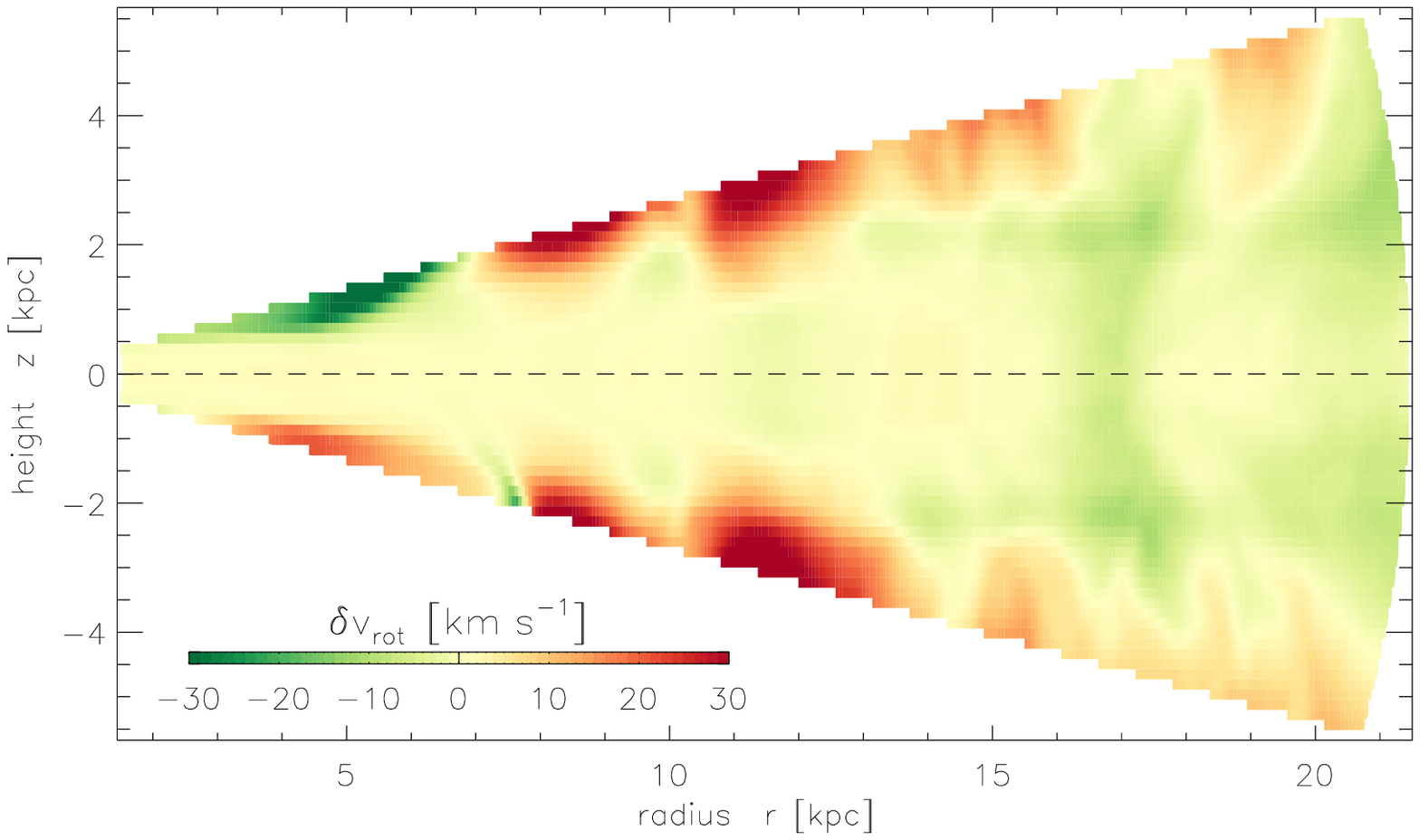}
  \caption{Results from a dynamical dynamo simulation including
    mean-field effects. Colour-coded lag in the rotation velocity
    $\vv_\phi(r,z)$ averaged over azimuth. Values $|\delta \vv| >
    30\kms$ have been clipped for clarity. \emph{Top:} Predicted
    deviation from the initial rotation profile based on the effect of
    the Lorentz force. \emph{Bottom:} the actual deviation seen in the
    simulation, which appears more washed out owing to the effect of
    the turbulent viscosity but otherwise agrees markedly well
    (see also the lower panel of Fig.~9 in
    Gressel et al. 2013
    for the actual field distribution).}
  \label{fig:sim}
\end{figure}

The deviation in rotation after the magnetic field has grown into
saturation is shown in the lower panel of Fig.~\ref{fig:sim}. Here and
in the following figure, we plot azimuthally averaged rotation
velocities. The colour coding is such that super-rotation is indicated
by red hues, sub-rotation corresponds to green hues, and neutral
rotation appears in yellow. For reasons of clarity, the data range has
been limited to $\pm 30\kms$, with values exceeding this level being
confined to high-latitude regions near the domain
boundary.
\footnote{The departures there are related to strong vertical
  fields in the disc halo, and simulations with an extended domain
  should be performed to exclude a spurious origin related to the
  chosen boundary conditions.}
Apart from being of more diffuse
appearance, the structure seen in this panel agrees very well with the
prediction obtained via Eq.~(\ref{eq3}), which is displayed in the
upper panel of the same figure, and which is based on computing
$\vv_{\rm mag}^2$ from the Lorentz force encountered in the
simulation. It is thus fair to say that the resulting deviations of
the rotational velocities in our MHD model can be explained by the
acting of magnetic forces.

The most pronounced difference between the two panels of
Fig.~\ref{fig:sim} is that the effect of the Lorentz force (seen in
the upper panel) appears negligible in the disc midplane, something
that is owed to the higher gas density there and which reduces the
relative importance of magnetic forces. This trend is moreover
reinforced when going radially inward, where the disc surface density
is higher. In contrast to this, the actual rotation profile seen in
the simulations shows some local deviations even in the midplane. This
may be explained by a coupling of overlying magnetically-affected
layers via a viscous drag force mediated by the turbulent viscosity
caused by SNe. A potential example for this can be seen around
$r=17\kpc$, where the rotation appears to be coupled over an extended
vertical patch of the disc.

\begin{figure}
  \includegraphics[width=\columnwidth]{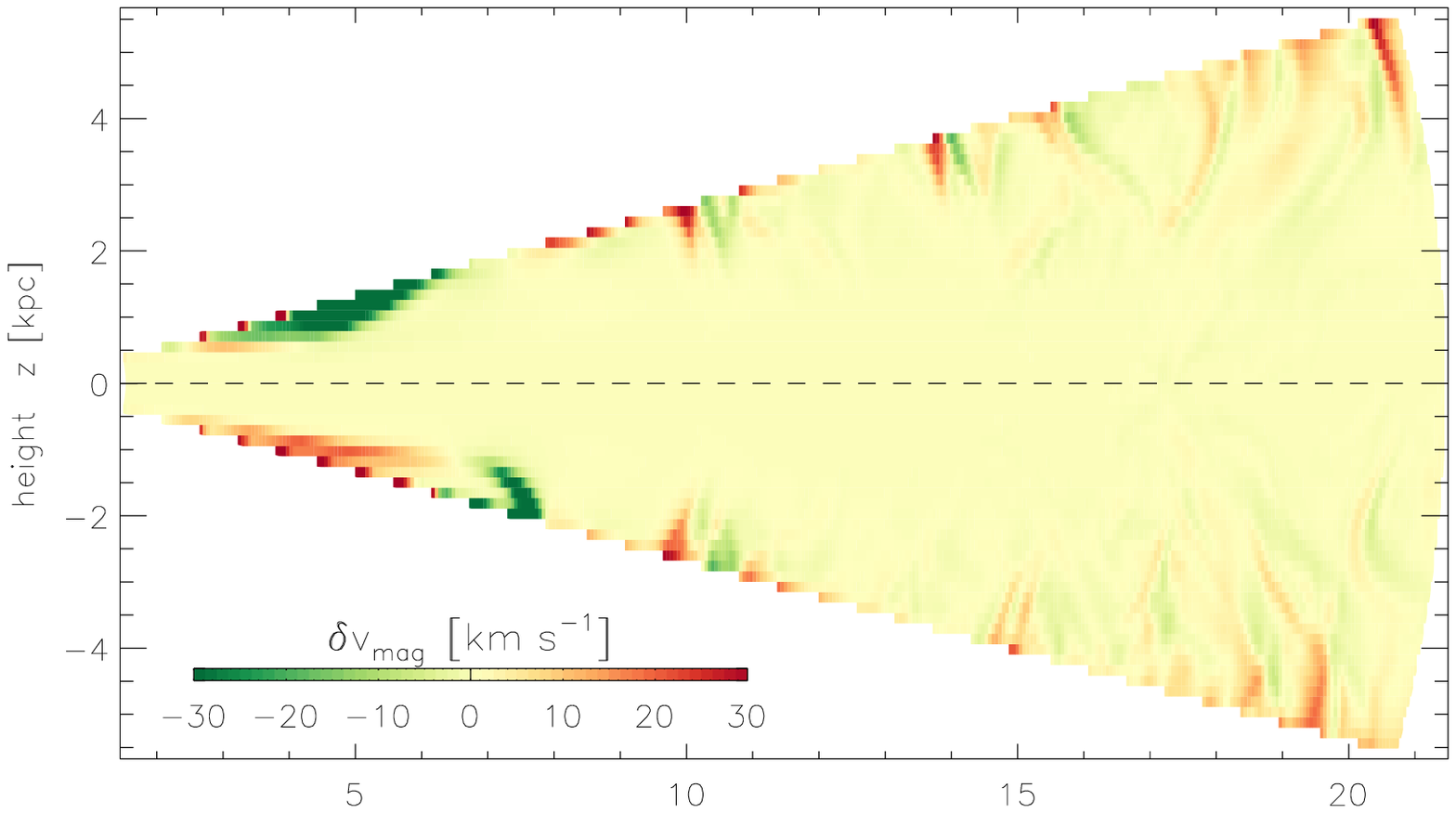}\\[6pt]
  \includegraphics[width=\columnwidth]{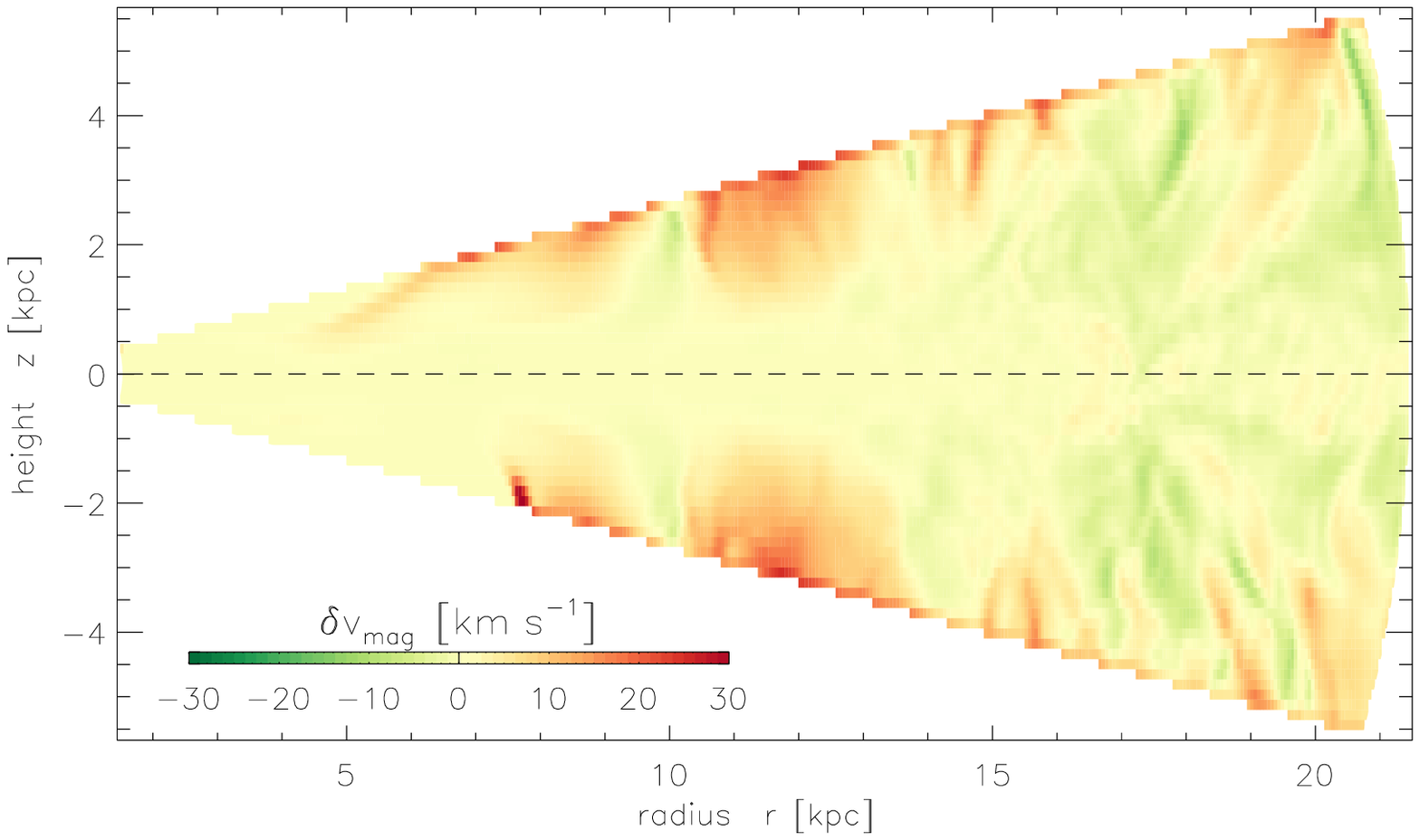}
  \caption{Same as the upper panel in Fig.~\ref{fig:sim} above, but
    comparing the contributions from different field
    components. \emph{Top:} effect of the poloidal field
    $B_z(r,z)\,\zz+B_r(r,z)\,\rr$, which predominantly affects
    high-latitude layers presumably via magnetic pressure
    gradients. \emph{Bottom:} effect of the azimuthal field
    $B_\varphi(r,z)$ component, contributing to super-rotation outside
    the reversal at $r=10\kpc$ and providing weak pressure-support in
    the outer disc away from the midplane.}
  \label{fig:sim2}
\end{figure}

To facilitate a more detailed interpretation of the various features
seen in the rotation cross-section of the disc, we now turn to
Fig.~\ref{fig:sim2}, where we plot the contributions to $\vv_{\rm
  mag}^2$ stemming from the poloidal field $B_z\,\zz+B_r\,\rr$ (upper
panel), and the azimuthal field component $B_\varphi$ (lower
panel). It can be seen that the poloidal field only affects the very
upper layers of the disc. Given the limited vertical extend of our
model, we should express some caution in interpreting these features
as they may be partly affected by the chosen boundary conditions.
Still, the characteristic patches seen near the field reversal
at $r=10\kpc$ actually serve as a nice example that vertical fields
enter the force balance via magnetic pressure gradients. Here, a
localised region of strong vertical field leads to a pressure support
(and hence sub-rotation, green) on its outside, and the opposite
effect (red) on the inside. Similar effects are seen farther out in
the disc but show a more complex structure. These features outside
$10\kpc$ may be related to the Parker instability also seen in
the simulations of \cite{2013ApJ...764...81M}.

As can be seen in the lower panel of Fig.~\ref{fig:sim2}, the more
large-scale variations in $\vv_{\rm rot}$ are clearly caused by the
azimuthal disc field $B_\varphi$.  Only the asymmetric rotation of the
upper and lower disc inside $6\kpc$ is due to the combined
symmetric radial component of the dynamo and a vertical field
going through the disc. This mixed parity of the field is the reason
for the antisymmetric change of the rotational velocity arising from
the term $B_z \partial_z B_r$ in Eq.~\ref{eq1}.  The strongest effect
causing super-rotation (red) is related to the dynamo mode with A0
parity residing between $8\kpc$ and $10\kpc$, and the combined dynamo and
MRI mode located between $10\kpc$ and $13\kpc$. At the interfaces between
these modes, the azimuthal field $B_\varphi$ vanishes, which leads to
local gradients in the Lorentz force. Both of the red regions have
green regions adjacent on the outside and the overall picture is again
consistent with the dominant effect stemming from magnetic
forces. Even though these features appear fairly washed-out in the
actual simulation, they are tied to particular field reversals and
their character is clearly local.

Because of the A0 parity, the super-rotation near the field reversal
(at $10\kpc$) does not affect the disc midplane. With the midplane
harbouring the peak in the gas density on one hand, and $B_\phi=0$ on
the other hand, this does seem a little surprising. In contrast to the
picture based on the ${\,\mathbf j}\times{\mathbf B}$ alone, the
simulation (cf. the lower panel of Fig.~\ref{fig:sim}) does show some
weak sub-rotation around $(r,z)=(12,0)\kpc$. Given the long-wavelength
vertical modulation, one may be tempted to interpret these
fluctuations as low-wavenumber MRI channels, where the effect of
field-line tension is complemented by inertial forces. The global
tendency of sub-rotation around the midplane outside $10\kpc$ is
consistent with the scale length of $10\kpc$ for the exponential fit
to the magnetic field strength in the model.

The variation of the rotation of the Milky Way, as is shown in
Fig.~2 of \cite{2012ApJ...755L..23R}, may be due to a field reversal
between $10\kpc$ and $15\kpc$, mainly caused by the azimuthal field, although
the details of the reversal differ in our model.  We should remark
that the magnetic field strength of the dynamo may still be
underestimated and the contribution of the small-scale turbulent field
is neglected in this type of model. Therefore one could have slightly
stronger changes of the rotation by the action of the Lorentz force in
real galaxies.


\section{Discussion}

From Eq.~(\ref{eq5}) we obtain the effect of the Lorentz force from
the total field strength, supposing we know the fraction of the
different components and the vertical field on the surface of the
disc. For the regular or an anisotropic field, we can estimate the
fraction of radial to azimuthal field by the pitch angle.  The vertical
component of the regular field  near the disc plane is
negligible, as seen from polarised emission of edge-on galaxies
(e.g. \citet{2011A&A...531A.127S}). The surface term ($B_r(h)B_z(h)$) reduces
the magnetic velocity for observed X-shaped geometries. For a
typical pitch angle of 20$^\circ$, the field contributing to the
polarised emission only acts like an azimuthal field. The
other part of the field contributing to the total intensity will act
like isotropic turbulence. Therefore the profile of the polarised intensity
should be flatter than $r^{-2}$ for  a centripetal force action.

It is observationally unclear how far one can extrapolate the
exponential law of the magnetic field strength and how reliable the
assumption of energy equipartition between magnetic field and cosmic rays
is in the outer disc.  In the case of a valid exponential law
of the regular magnetic field in the outer disc, a reduction of the
circular velocity occurs outside the scale length, that is, where the
tension force is dominated by the pressure force of the azimuthal
field. Apart from ram pressure stripping, this may be another source for truncations
of the gaseous disc. Dynamical simulations of ram pressure stripping suggest this
\citep{2014ApJ...784...75R}.  Stars formed within the slower rotating
gas will decouple from the magnetic force and migrate inwards.

With the present data we do not find any support 
for the possibility of attributing flat
rotation curves to magnetic forces. This has been claimed previously
with the intention of reducing the required dark matter content of
galaxies. On the contrary, at large radii the observed flat gas rotation
may demand even more extended
dark matter profiles because of an additional magnetic pressure support.
This is, however, speculative because the observations of extended disc
rotation are mainly done for extended gaseous discs, where the Alfv\'en
velocity is still small compared to the rotation velocity.  Additional
direct measurements of radio emission in the outer discs in the low-frequency range or
 of Faraday rotation of background radio sources \citep{2013AN....334..548B} will
give better constraints than our simple extrapolation of the scale
length.

Our dynamo model exhibits nearly no influence of the magnetic field on
the rotation in the disc, with a weak sub-rotation outside the radial
scale length of the field.  Outside the midplane of the disc,
variations of the order of 10\% of the rotation velocity may be caused by magnetic field
reversals.  The super-rotation appearing at the outer vertical
boundaries caused by the flat radial profile of the azimuthal field may be due
to boundary effects and no proper modelling of the halo gas.  Another
interesting feature seen in our dynamo model is the antisymmetric
disturbance of the circular velocity away from the midplane by a
symmetric disc field, together with a vertical field threading the
disc.

\section{Summary}

We estimated the effects of the Lorentz force onto the gas rotation of galaxies assuming
equipartition of the magnetic field with cosmic ray energy.

The scale length of the regular magnetic field is rather large compared to the optical disc size.
The scale for the gas density varies between a quarter and a half of the magnetic scale length.
This translates into a constant or slowly growing Alfv\'en velocity along the disc.
Stripped galaxies may be an exception.

An exponential distribution of the magnetic field leads always to a centrifugal force (reduction of
the rotational velocity) outside one exponential scale length.
The centripetal force (increase of the rotational velocity) inside  one exponential scale length is too weak 
because of the low Alfv\'en velocity in the disc.
A strong turbulent field in the inner region will further reduce the centripetal force.
Reversals of the large scale magnetic field on scales below $1\kpc$ may cause variations
of the rotation velocity.

Recent dynamic dynamo models confirm the above findings, independent of the equipartition argument.

\begin{acknowledgements}
  Part of this work was supported by the German \emph{Deut\-sche
    For\-schungs\-ge\-mein\-schaft (DFG)\/}, project numbers FOR1254
  and FG1257. We thank Marita Krause for carefully reading the manuscript
  and the referee Eduardo Battaner for useful comments.
\end{acknowledgements}

\bibliographystyle{aa}
\bibliography{refs}

\end{document}